\documentclass[runningheads]{llncs}

\usepackage{floatrow}
\newfloatcommand{capbtabbox}{table}[][\FBwidth]

\usepackage{todonotes}
\usepackage{fixltx2e}
\usepackage{amssymb}
\usepackage{graphicx}
\usepackage{wrapfig}
\usepackage{subfig}
\usepackage{subfloat}
\usepackage{url}
\usepackage{amsmath}
\usepackage{verbatim}
\usepackage{multirow}
\usepackage{paralist}
\usepackage{mathptmx}
\usepackage{times}
\usepackage{dirtree}
\usepackage{tabularx}
\usepackage{tikz}
\usetikzlibrary{arrows,%
                calc,%
                chains,%
                decorations.pathmorphing,%
                decorations.pathreplacing,%
                decorations.shapes,%
                graphs,%
                matrix,%
                positioning,%
                shapes.geometric,%
                shapes,%
                shapes.symbols,%
                trees
   			  }

\usepackage{graphicx}
\usepackage{pgfplots}
\usepackage{floatrow}

\usetikzlibrary{automata,arrows}
\usetikzlibrary{datavisualization}
\usetikzlibrary{datavisualization.formats.functions}
\usetikzlibrary{positioning, calc, backgrounds, arrows}
\usetikzlibrary{shapes.geometric}
\usetikzlibrary{decorations.pathreplacing}
\usetikzlibrary{patterns}

\let\olddefinition\definition
\def\definition{\olddefinition\footnotesize}

\newcommand{\optional}[1]{}

\usepackage[ruled,vlined,boxed,linesnumbered]{algorithm2e}
\SetKwRepeat{Do}{do}{while}
\SetKwInOut{Input}{input}
\SetKwInOut{Output}{output}

\graphicspath{{images/}}

\newcommand{\event}{e}
\newcommand{\eventid}{e_{\mid c}}
\newcommand{\eventa}{e_{\mid a}}
\newcommand{\eventime}{e_{\mid t}}

\newcommand{\trace}{\tau}
\newcommand{\elog}{\mathcal{L}}
\newcommand{\activities}{\mathcal{A}}
\newcommand{\timestamps}{\mathcal{T}}
\newcommand{\cc}{C_c}
\newcommand{\ac}{C_a}
\newcommand{\deltas}{\mathcal{D}}

\newcommand{\cIDs}{\mathcal{C}}
\newcommand{\aIDs}{\mathcal{V}}
\newcommand{\obs}{\mathcal{O}}

\begin{document}

\title{Efficient Checking of Temporal Compliance Rules Over Business Process Event Logs}
\author{	Adriano Augusto\inst{1} \and
		    Ahmed Awad\inst{2} \and
			Marlon Dumas\inst{2}
			 }
\titlerunning{ }
\authorrunning{ }

\institute{
           University of Melbourne, Australia\\
   		    a.augusto@unimelb.edu.au
           \and
           University of Tartu, Estonia\\
           \{ahmed.awad, marlon.dumas\}@ut.ee
}

\maketitle \addtocounter{footnote}{-2}

\begin{abstract}
Verifying temporal compliance rules, such as a rule stating that an inquiry must be answered within a time limit, is a recurrent operation in the realm of business process compliance. In this setting, a typical use case is one where a manager seeks to retrieve all cases where a temporal rule is violated, given an event log recording the execution of a process over a time period. Existing approaches for checking temporal rules require a full scan of the log. Such approaches are unsuitable for interactive use when the log is large and the set of compliance rules is evolving. This paper proposes an approach to evaluate temporal compliance rules in sublinear time by pre-computing a data structure that summarizes the temporal relations between activities in a log. The approach caters for a wide range of temporal compliance patterns and supports incremental updates. Our evaluation on twenty real-life logs shows that our data structure allows for real-time checking of a large set of compliance rules.
\end{abstract}

\section{Introduction}\label{sec:intro}
Enterprise information systems maintain detailed records of activity executions across the business processes they support. These records can be extracted in the form of \emph{event logs}, which enable a range of analytics capabilities, collectively known as \emph{process mining}. A common class of process mining techniques is \emph{conformance checking}~\cite{conformanceCheckingBook18}, where process executions are compared against a reference model or a set of rules, to detect and to analyse deviations. 
Within this field, a common operation consists in identifying process instances (cases) that violate one or more compliance rules, a.k.a.\ \emph{(business process) compliance checking}~\cite{complianceRequirements2019}.\footnote{This operation is equivalent to that of detecting deviations between process executions and \emph{declarative process models}, i.e.\ models that capture rule-sets diagrammatically~\cite{CMF15,multiPerspectiveDECLARE16}.}
Previous studies have tackled the problem of compliance checking w.r.t.\ various types of rules, categorized into compliance patterns in~\cite{elgammal2016formalizing,complianceRequirements2019}.
Within this framework, the goal of compliance checking is to detect and analyse deviations between the process executions recorded in an event log, and a set of compliance rules, each of which is an instance of a compliance pattern. 


This paper addresses the problem of efficiently checking a class of rules known as \emph{temporal compliance rules}, which capture constraints on the time at which the events in a process execution may or should occur. Existing approaches for checking such rules require a full scan of the log. These approaches are unsuitable for interactive use when the log is large and the set of compliance rules is continuously evolving. 
To tackle this limitation, this paper puts forward a data structure that succinctly describes the temporal relations between activity instances in an event log. This data structure can be queried to identify non-compliant cases in a sublinear time to the event log size. The proposed approach is flexible enough to handle a wide range of temporal compliance patterns. The paper reports on an empirical evaluation on 20 real-life logs aimed at validating the applicability of the approach for interactive compliance checking.





\section{Background and Related Work}\label{sec:background}

In this section, we introduce the concepts of event log and trace, we give an overview of process compliance patterns, and we discuss the research gap addressed in this paper.

\subsection{Preliminaries}\label{sec:preliminaries}


\begin{definition}\textbf{Event --}\label{def:event}
An \emph{event} $\event$ can be defined as a tuple $(x_1, x_2, \dots, x_n)$, where each element $x_i$ captures an attribute of the event, and at least three attributes are present: the label identifying a specific process instance ($c$ -- event case label); the label of the activity the event refers to ($a$ -- event activity label); and the timestamp ($t$ -- event timestamp). Additional attributes can capture the process resource who executed the activity, customer information, etc. 
\end{definition}

In the following, given an event $\event$, we refer to its three required attributes with the notation $\eventid$, $\eventa$, $\eventime$. It is well-known that some systems record an event when an activity is started and an event when an activity is completed. Although this would not affect our approach, it would unnecessarily complicate its explanation. Hence, we will cover this case separately (in Section~\ref{sec:itupdate}) and, for the reminder, we assume that the event timestamp ($\eventime$) refers only to the completion time of the event activity ($\eventa$). 
\begin{definition}\textbf{Event Log --}\label{def:log}
An \emph{event log} $\elog$ is a sequence of events $\langle \event_1, \event_2, \dots, \event_n \rangle$, such that all the events are ordered by their timestamp. Formally, $\forall \event_i \in \elog \mid i \in [1, n-1] \cap \mathbb{N} \Rightarrow {\event_i}_{\mid t} \leq {\event_{i+1}}_{\mid t}$.
\end{definition}

We note that events recorded in an event log may not be ordered by timestamp; timestamp errors are a well-known data quality issue~\cite{suriadi2017event}. However, reordering the events by timestamp is a trivial activity, and there exist automated techniques for fixing timestamp granularity~\cite{conforti2020automatic}. Hence, we assume the event log is a sequence of timestamp-ordered events, and our approach operates on this assumption.


\begin{definition}\textbf{Trace (Case) --}\label{def:trace}
Given an event log $\elog$, a \emph{trace} of the event log $\trace \in \elog$ is a sequence of events, $\trace = \langle \event_1, \event_2, \dots, \event_n \rangle$, such that: i) all the events belong to the event log; ii) all the events are ordered by their timestamp; and iii) all the events have the same case label attribute. Formally, $\forall \event_i \in \trace \mid i \in [1, n-1] \cap \mathbb{N} \Rightarrow {\event_i}_{\mid c} = {\event_{i+1}}_{\mid c} \wedge {\event_i}_{\mid t} \leq {\event_{i+1}}_{\mid t}$. A trace is effectively a process execution instance, hence, it can also be referred to as a \emph{case}.
\end{definition}

\subsection{Compliance Patterns}\label{sec:compliance:patterns}

\begin{wrapfigure}[12]{R}{0.60\textwidth}
    \centering
    \includegraphics[scale=0.59]{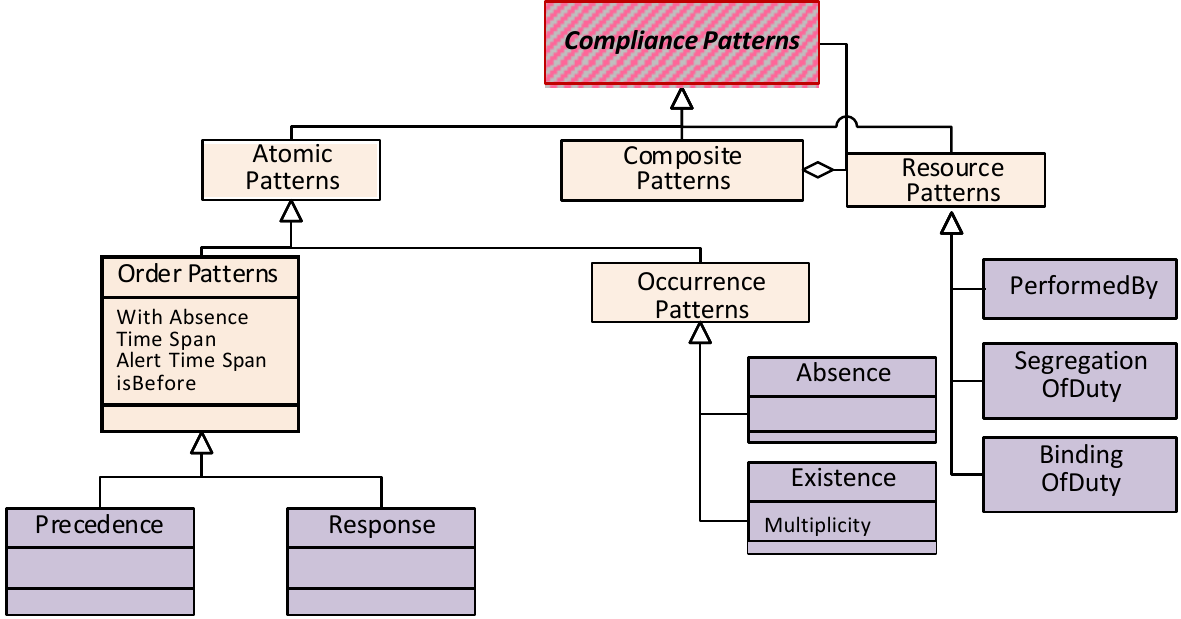}
    \caption{Categorization of compliance patterns}
    \label{fig:compliance:patterns}
\end{wrapfigure}

In traditional conformance checking~\cite{conformanceCheckingBook18}, a trace is compared against a process model to quantify the deviation between the prescribed behaviour (the model) and the observed behaviour (the trace). Mostly, conformance checking relates to control flow aspects of the process execution, although multi-perspective approaches have been developed~\cite{conformanceCheckingSurvey19}. However, these techniques require the process models to be enriched with resource, temporal, and data constraints to compute deviations. Moreover, identified deviations are holistic on an end-to-end trace level. In many cases, it might be required to check deviations on a much finer granularity, e.g., on the level of activities, e.g., absence, existence, or pairs of activities, e.g., co-existence, mutual exclusion, and temporal and resource variants thereof. Such finer granularity checks are referred to as compliance checking, and \emph{compliance patterns} are used for categorizing the type of compliance requirements~\cite{elgammal2016formalizing,CMF15,complianceRequirements2019}. Fig.~\ref{fig:compliance:patterns} shows a classification of patterns for business process compliance as per~\cite{elgammal2016formalizing}. 

Occurrence patterns are concerned with activity having been executed (\emph{Existence}) or not (\emph{Absence}) within a process instance. Order patterns are concerned with the execution order between pairs of activities. The \emph{Response} pattern (e.g., \emph{Response(A, B)}) states that if the execution of activity $A$ is observed at some point in a process instance, the execution of activity $B$ must be observed in some future point of the same case before its termination. A temporal window can further restrict these patterns. For instance, we need to observe $B$ after $A$ in no more than a certain amount of time units. Alternatively, we need to observe $B$ after observing $A$, where at least a certain amount of time units have elapsed. Definition~\ref{def:response} formalizes the \emph{Response} pattern. Note that we get the unrestricted form of the Response pattern by setting $\Delta t$ to a sufficiently large value and $\theta$ to $\leq$. That is, $B$ has to \emph{eventually} be observed after $A$ with no further restrictions on the time window.

\begin{definition}\textbf{Response --}\label{def:response}
Given two activities $A$ and $B$ and a trace $\trace = \langle \event_1, \event_2, \dots, \event_n \rangle$, $\trace \in \elog$, we say that $\trace \models Response(A,B, \Delta t, \theta)$ if and only if $\forall e_i \in \trace: {e_i}_{\mid a}=A~\exists~e_j: {e_j}_{\mid a}=B \wedge {e_i}_{\mid t} \leq {e_j}_{\mid t} \wedge ({e_j}_{\mid t} - {e_i}_{\mid t})~ \theta~ \Delta t$, where $\Delta t$ represents the time window between $A$ and $B$, and $\theta$ represents when (i.e., after, before, or exactly at) we expect the observation of $B$ after $A$ with respect to $\Delta t$.
\end{definition}

Conversely, the pattern \emph{Precedes(A,B)} states that if the execution of activity $B$ is observed at some point in the trace, $A$ must have been observed before (Definition~\ref{def:precedes}). 

\begin{definition}\textbf{Precedes --}\label{def:precedes}
Given two activities $A$ and $B$ and  a trace $\trace = \langle \event_1, \event_2, \dots, \event_n \rangle$, $\trace \in \elog$, we say that $\trace \models Precedes(A,B, \Delta t, \theta)$ if and only if $\forall e_i \in \trace: {e_i}_{\mid a}=B~\exists~e_j: {e_j}_{\mid a}=A \wedge {e_j}_{\mid t} \leq {e_i}_{\mid t} \wedge ({e_j}_{\mid t} - {e_i}_{\mid t})~ \theta~ \Delta t$.  where $\Delta t$ represents the time window between $A$ and $B$ and $\theta$ is how we expect the observation of $B$ after $A$ with respect to $\Delta t$.
\end{definition}

 Resource patterns are concerned with constraints on performers of pairs of activities, such as separation of duty or bind of duty. Finally, such patterns can be combined to form composite patterns using different types of logical operators. In this paper, we focus on occurrence and order patterns and their temporal variants. 

\subsection{Related Work}
\label{sec:related:work}

The literature about compliance checking of business processes is vast. For our purposes, we focus on compliance checking over event logs; we refer to them as auditing. For a more comprehensive survey, we refer the reader to~\cite{surveyComplianceChecking18}.

We can generally categorize auditing based on the process perspective, i.e., control-flow, data, resources, or temporal. Additionally, we can divide these categories based on the underlying formalism and technology. One of the earliest works on compliance auditing is by Agrawal et al.~\cite{tamingComplianceDB06}, where process execution data is loaded into relational databases, and compliance is checked by identifying anomalous behaviour. For this, so-called workflow graphs of the rules and the process are compared, looking for deviations. The approach covers control-flow-related aspects.

Model checking techniques have been proposed to validate process logs against control-flow and resource-aware compliance rules~\cite{processMiningVerificationLTL05}.
Adapting conformance checking techniques for auditing purposes has been proposed in~\cite{auditing2.010}. Ramezani et al.~\cite{complianceCheckingAlignment12,complianceCheckingTemporalRequirements13} propose alignment-based detection of compliance violations for control-flow and temporal rules. In~\cite{complianceCheckingResourceData14} another alignment-based approach for resource-related compliance violations is presented. 

%
De Murillas et al.~\cite{connectingDBPM19} propose a metamodel and a toolset to extract process-related data from logs of operational systems, e.g., relational databases, and populate their metamodel. The metamodel is stored in a relational database. The authors show how different queries can be translated to SQL. Using relational databases provides support for a wide range of queries against process data. However, such queries are complex (using nesting, joins, unions), and the underlying data cannot be entirely kept in memory. OLAP-like analysis of process data is another option for resolving queries following slice, dice, drill-down and roll-up operators of data cubes; hence, several approaches have been proposed to store event data in so-called process cubes~\cite{relationalDWPM15,multiDimensionalPM15}. Relational databases have also been used for declarative process mining~\cite{declarativeMiningSQL16}, which can be seen as an option for checking logs against compliance rules.

The compliance patterns can be checked by the ANSI SQL operator \texttt{Match\_Recognize} (MR). In essence, MR verifies patterns as regular expressions, where tuples of a table are the symbols of the string to search for matches within. MR runs linearly in the number of tuples of the table. In our case, the tuples are the events in the log. In practice, the operational time can be enhanced by paralleling the processing, e.g., partitioning the tuples by the case identifier. Nevertheless, this does not change the linearity of the match concerning the number of tuples in the table. Recently, there has been work to further speed up MR by utilizing indexes in relational databases~\cite{indexAcceleratedPatternMatching21} for strict contiguity patterns, i.e., patterns where events are in strict sequence. Order compliance patterns frequently refer to eventuality rather than strict order, limiting the use of indexes to accelerate the matching process.


Storing and querying event logs using graph databases has also been investigated. Esser et al.~\cite{multiDimlEventGrahDB21} provide a rich multidimensional model for event data using labelled property graphs realized on top Neo4J. The authors show how their model supports several classes of queries on event data. All compliance patterns can be represented as queries against their model.

In summary, all the discussed techniques execute in linear time with respect to the number of events in the log. Moreover, each time a check is invoked, the whole log has to be processed. In many cases, several traces are irrelevant to the rule being checked or do not expose violations, resulting in wasting time with repeated checks.

\section{Approach}\label{sec:approach}
In this section, we describe our approach to generate, in \emph{linear} time on the number of events in a log, a data structure that can be used to check temporal compliance rules in \emph{sublinear} time on the number of traces in the given log.
First, we discuss how to generate and query our data structure. Then, we analyse its time and space complexity, comparing it with approaches based on \texttt{Match\_Recognize}. Lastly, we explain how our data structure allows for an efficient online update and discuss additional optimizations.

\subsection{Data Structure Components and Generation}

The design of a data structure for efficiently checking temporal compliance rules is a problem of striking a trade-off between time efficiency (i.e., how much time it takes to build, update, and access the data structure) and space efficiency (i.e., how much memory space it takes to store the data structure). 

Given an event log $\elog$, the information required to check temporal compliance rules (discussed in Section~\ref{sec:querying}) is the following.
For each pair of events $(e_i, e_j) \in \elog$ such that ${e_i}_{\mid c} = {e_j}_{\mid c} \wedge i < j$, we need to store or efficiently access/calculate:
i) the time elapsed between $e_i$ and $e_j$: $\Delta t_{(e_i, e_j)} = {e_j}_{\mid t} - {e_i}_{\mid t}$;
ii) the case labels where we observed $e_i$;
iii) the case labels where we observed $e_j$.

When an event log is stored in memory as is described in Definition~\ref{def:log}, to check a temporal compliance rule it is necessary to calculate these three pieces of information in run-time. While such an approach may be space-efficient since it requires only storing the original log, it is clearly not time efficient -- we would need to read the entire log each time we would like to check a set of temporal compliance rules. On the opposite side of the solutions spectrum, there would be a special-purpose data structure that stores all three pieces of information in memory and (ideally) allows for checking temporal compliance rules in constant time. This latter solution would clearly be time-efficient, but it may theoretically compromise space efficiency -- storing information i) would require $e(e+1)/2$ memory cells (in the worst case).~\footnote{This is equivalent to $10^{11}$ memory cells, for a log of one million events.}

Considering today's enterprise IT architecture and capabilities, we designed a data structure intending to optimize time efficiency rather than space efficiency, given that memory is a low-cost staple and compliance checking can be a real-time problem.

Our data structure can capture all the required information to check temporal compliance rules, while reducing information redundancy and optimizing access time. Our data structure is based on six maps and two integer counters. Table~\ref{tab:datastruct} shows an overview of the data elements composing our data structure. 
\begin{table}
\centering
{\scriptsize{
\def\arraystretch{1.5} 
\begin{tabular}{l|c|p{0.83\linewidth}}
	
    \hline
	\textbf{Type}
	& \textbf{Symbol}
	& \textbf{Description}
    \\\hline
	
    Integer & $\cc$ & Counter to facilitate the mapping of event case labels to integers\\
    
    Integer & $\ac$ & Counter to facilitate the mapping of event activity labels to integers\\
    
    Map & $\cIDs$ & Maps an event case label to its integer ID\\
    
    Map & $\aIDs$ & Maps an event activity label to its integer ID\\
    
    Map & $\obs$ &  Maps an activity integer ID to the set of case integer IDs where that activity was observed\\
    
    Map & $\activities$ & Maps a case integer ID to the (ordered) list of activity integer IDs of the events having that case ID\\
    
    Map & $\timestamps$ & Maps a case integer ID to the (ordered) list of timestamps of the events having that case ID\\
    
    Map & $\deltas$ & Maps a pair of activity integer IDs to a red-black tree with nodes ($\Delta t$, case integer ID), ordered by $\Delta t$\\\hline

	\end{tabular}
}}
  	\caption{Our data structure elements.}\label{tab:datastruct}
\end{table}
\begin{figure}[htb]
  \centering 
  \includegraphics[width=0.89\textwidth]{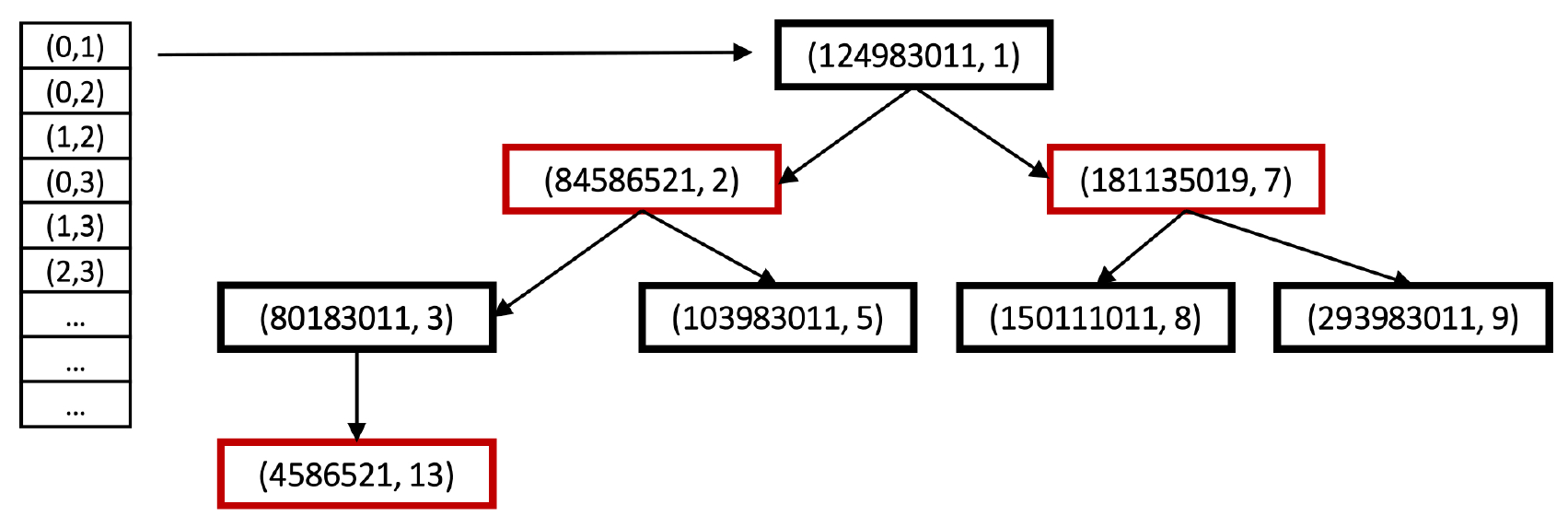} 
  \caption{Graphical representation of the map $\deltas$.}
  \label{fig:deltas}
\end{figure}
\begin{figure}[htb]
  \centering 
  \includegraphics[width=0.95\textwidth]{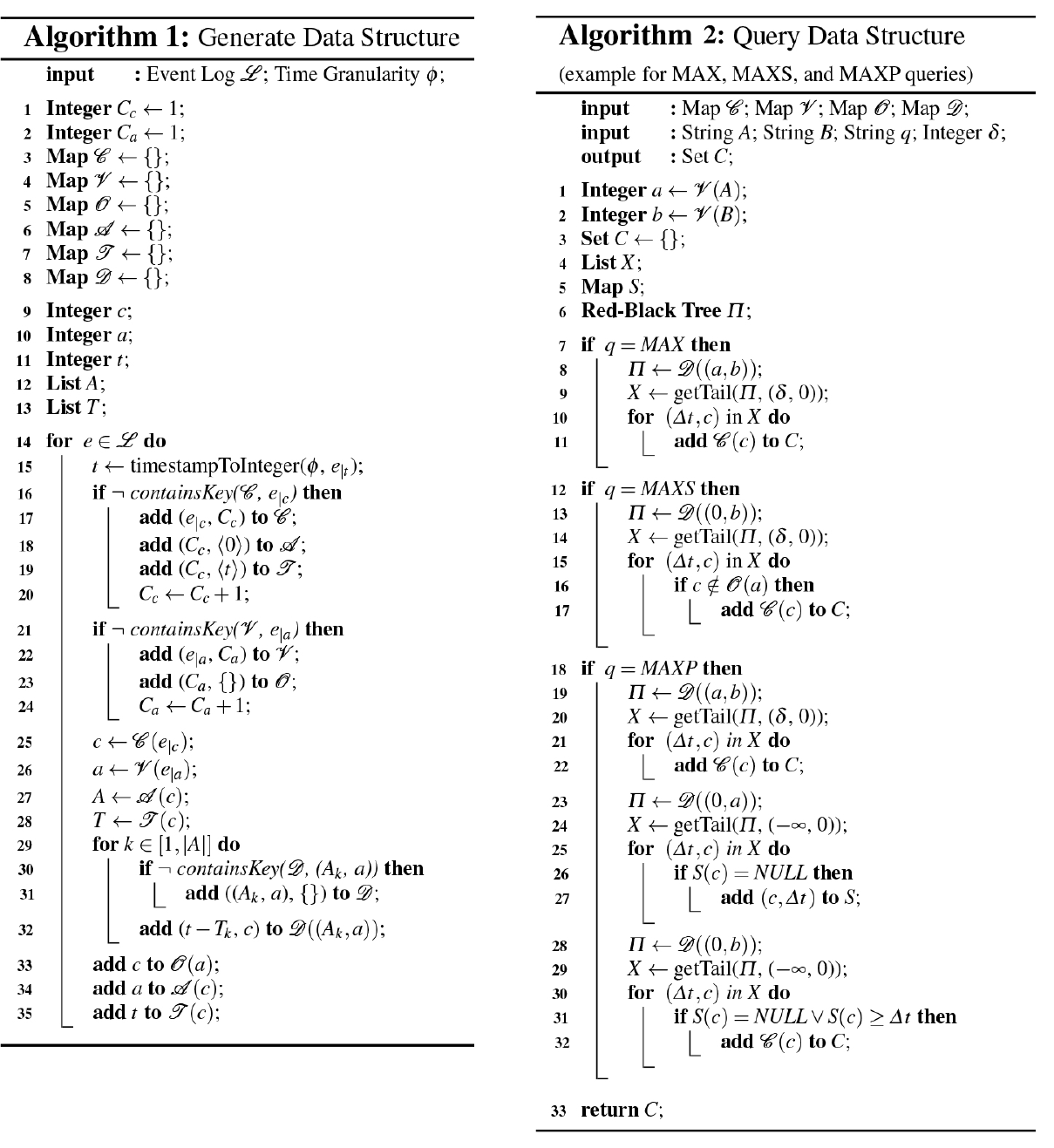} 
  \vspace{-5mm}
  \label{fig:algo}
\end{figure}
The counters $\cc$ and $\ac$ facilitate the conversion of case and activity labels into integers, this is mostly required to reduce memory space~\footnote{Case and activity labels can have more than 20 alphanumeric characters, i.e., 20+ bytes.}, but it can also reduce the access time complexity to maps.~\footnote{Maps with string keys have access time complexity proportional to their keys' average length.} The maps $\cIDs$ and $\aIDs$ respectively match case and activity labels to their integer IDs, e.g., an element of $\aIDs$ is a pair (activity label, activity integer ID). In the following, we refer to case and activity integer IDs as if they were the original case and activity labels. The map $\obs$ keeps track of the cases where we observed a specific activity. An element of $\obs$ is a pair (activity integer ID, set of integer cases IDs). The maps $\activities$ and $\timestamps$ are effectively a compact representation of the event log. Both maps have case integer IDs as keys, given a key $c$, $\activities(c)$ is the list of activity integer IDs corresponding to the timestamp-ordered sequence of activity labels of the events (observed in the original log) having the case label whose integer ID is equal to $c$. While, $\timestamps(c)$ is the list of ordered timestamps (in integer form) of the events (observed in the original log) having the case label whose integer ID is equal to $c$. The maps $\activities$ and $\timestamps$ are required only to facilitate the update of the data structure, in the case of incomplete traces (discussed in Section~\ref{sec:itupdate}).

Lastly, the core element of our data structure is the map $\deltas$. This map stores a compact representation of the $\Delta t_{(e_i, e_j)}$ for all $e_i, e_j$ in a given log, such that ${e_i}_{\mid c} = {e_j}_{\mid c} \wedge i < j$. Fig.~\ref{fig:deltas} shows a graphical representation of the map $\deltas$. The keys of $\deltas$ are pairs of activity integer IDs. The values of $\deltas$ are red-black trees~\cite{cormen2009introduction}, where each tree node is a pair ($\Delta t_{(e_i, e_j)}$, case integer ID), and the order is based on the $\Delta t_{(e_i, e_j)}$ (using the case integer ID as tiebreaker).
Red-black trees are self-balancing binary search trees that provide efficient access, with a worst-case time complexity of $O(log(n))$ (n = number nodes) and $O(1)$ on average (amortized).
We note that $\Delta t_{(e_i, e_j)}$ are expressed in the form of integers capturing the minutes, hours, or days elapsed between the two events $e_i$ and $e_j$. 

We now turn our attention to Algorithm~1, which describes how to populate our data structure. Alg.~1 takes two inputs: the event log; and a parameter to determine the granularity of the timestamps when converted to integers (i.e., minutes, hours, days).
The algorithm begins by initializing all the data structure elements (lines 1-8) and a set of temporary variables (lines 9-13). Then, it iterates over all the events recorded in the event log (line 14), and for each event, it does the following. 

The event timestamp ($\eventime$) is retrieved and converted into an integer. If the event case label ($\eventid$) has never been observed before, a new case integer ID is generated, and new lists are created for storing activity integers ID and timestamps for the events of that case (lines 16-20). These two newly created lists are initialised with one element each: $0$  -- an integer referring to an artificial start activity; and $t$ -- the $\eventime$ in its integer form. 
Similarly, if the event activity label ($\eventa$) has never been observed before, a new activity integer ID is generated, and a new set is created for storing the cases where that activity will be observed (lines 21-24).

Next, the temporary variables are initialised (lines 25-28): $c$ -- the current event case integer ID; $a$ -- the current event activity integer ID; $A$ -- the list of activity integer IDs previously observed for the current event case integer ID; $T$ -- the list of integer timestamps of the previously observed activity integer IDs for the current event case integer ID.

At this point, the algorithm can update the map $\deltas$. For each activity integer ID in the list $A$ ($A_k$, line 29), if the pair of activity integer IDs ($A_k$, $a$) has never been observed before, the pair is added as a key in $\deltas$, with an empty tree as its current value (line 31). Lastly, the pair ($t - T_k$, $c$) is added to the tree associated to the pair ($A_k$, $a$), line 32. 

After updating $\deltas$, the algorithm updates the values of the maps $\obs$, $\activities$, and $\timestamps$ by recording the observation of the activity with integer ID $a$, in the case with integer ID $c$, having an integer timestamp $t$ (lines 33-35).

\subsection{Querying the data structure}\label{sec:querying}

Our data structure can be queried for checking a large set of the existing compliance patterns~\cite{elgammal2016formalizing}. For instance, the maps $\obs$, $\activities$, and $\timestamps$ can be queried to check \emph{occurrence} compliance patterns, such as 
\emph{existence} and \emph{absence}. 
While, the maps $\obs$ and $\deltas$ can be queried to check \emph{order} compliance patterns, such as \emph{leads to}, \emph{precedes}, and \emph{substitutes}, which can be combined with \emph{temporal} compliance patterns such as \emph{within}, \emph{after}, and \emph{exact}. Given that this study focuses on efficiently checking compliance rules that combine \emph{order} and \emph{temporal} patterns, we do not discuss the many additional compliance rules, the interested reader can refer to the literature~\cite{elgammal2016formalizing}. Table~\ref{tab:queries} shows a subset of rules that can be checked on our proposed data structure.
\begin{table}[t]
\centering
{\scriptsize{
\def\arraystretch{1.5} 
\setlength{\tabcolsep}{3pt} 
\begin{tabular}{l|l|p{0.65\linewidth}}
	
    \hline
    \textbf{Rule as in~\cite{elgammal2016formalizing}}
	& \textbf{Query Code}
	& \textbf{Query Output -- Given two activities $A$ (source) and $B$ (target)}
    \\\hline
    
    \emph{Within leads to} & MAX & \multirow{3}{1.00\linewidth}{Case labels where $B$ eventually follows $A$ after (MAX), or before (MIN), or exactly at (EXACT) $\delta$ time units.}\\
    
    \emph{At least after leads to} & MIN & \\
    
    \emph{Exactly at leads to} & EXACT & \\\hline
    
    \emph{Within substitute} &  MAXS & \multirow{3}{1.00\linewidth}{Case labels where $A$ is not observed and $B$ is observed after (MAXS), or before (MINS), or exactly at (EXACTS) $\delta$ time units from the starting time of the case.}\\
    
    \emph{At least after substitute} & MINS & \\
    
    \emph{Exactly at precedes} & EXACTS & \\\hline
    
    \emph{Within precedes} & MAXP & \multirow{3}{1.00\linewidth}{Case labels where $B$ eventually follows $A$ after (MAXP), or before (MINP), or exactly at (EXACTP) $\delta$ time units; \emph{and} case labels where $B$ is observed but not $A$; \emph{and} case labels where the first observation of $B$ occurs before the first observation of $A$.}\\
    
    \emph{At least after precedes} & MINP & \\
    
    \emph{Exactly at precedes} & EXACTP & \\\hline
	
	\end{tabular}
}}
  	\caption{Subset of compliance rules that can be checked on our proposed data structure.}\label{tab:queries}
\end{table}
Given two process activities $A$ and $B$, the rules with label MAX, MIN, and EXACT check for each case recorded in the event log where $B \leadsto A$ the validity of a constraint for the time elapsed between the completion of $A$ and the completion~\footnotemark~of $B$. The constraint can be a maximum (MAX), minimum (MIN), or exact (EXACT) amount of time units (e.g., hours, days).

The rules with label MAXS, MINS, and EXACTS check for each case recorded in the event log where $A$ is not observed the validity of a constraint for the time elapsed between the start of the case (i.e., process execution) and the completion~\footnotemark[\value{footnote}] of $B$. Similar to other rules, the constraint can be a maximum (MAX), minimum (MIN), or exact (EXACT) amount of time units.

The rules with labels MAXP, MINP, and EXACTP perform the same check of MAX, MIN, and EXACT, but, in addition, they also require $A$ to \emph{precede} $B$. Hence, these rules check also whether there exist cases where $B$ is not preceded by $A$.

\footnotetext{Or the start, depending on what activity lifecycle information is recorded in the event log.}

In general, unless a special-purpose data structure is used, in order to check these rules, one needs to iterate over the entire event log. In the worst case, the iteration would be one per rule to check. Our data structure allows for a more efficient querying approach. Alg.~2 describes how a query of type MAX, MAXS, or MAXP is resolved using our data structure. The inputs to Alg.~2 are four of our data structure maps ($\cIDs$, $\aIDs$, $\obs$, $\delta$), two activity labels ($A$, $B$), the query code ($q$), and the time constraint ($\delta$).
The activity labels are immediately converted into their corresponding integer IDs ($a$ and $b$, lines 1-2). Then, for the queries MAX and MAXP, the red-black tree of the pair ($a$, $b$) is retrieved (line 8 and 19), while for the query MAXS, the red-black tree of the pair ($0$, $b$) is retrieved (line 13).~\footnote{We recall that $0$ is the integer ID of the artificial start activity of any case, created during the generation of our data structure.}
We then retrieve all the ordered nodes of the red-black tree starting from the element that has $\Delta t \geq \delta$; we do this using the function \emph{getTail} (lines 9, 14, and 20).
For the queries MAX and MAXP, we extract from the tree nodes the case integer IDs and, after converting them into their original case label, we add them to the set $C$ (lines 10-11 and 21-22) -- the output of the query. Additionally, for the query MAXS, we filter the case integer IDs to retain only those cases where $a$ is not observed (lines 16-17).
While, for the query MAXP, we also retrieve the cases where we do not observe $a$ before $b$. To do so, for each case where $a$ is observed, we extract its lowest $\Delta t_(0, a)$ (lines 23-27). Then, from the cases where $b$ is observed, we retrieve those where $a$ was not observed or the $\Delta t_(0,b)$ is less than or equal to the lowest $\Delta t_(0, a)$ (lines 28-32).

Resolving the other queries listed in Table~\ref{tab:queries} only requires to retrieve different nodes from the red-black trees, precisely, those with $\Delta t \leq \delta$ (for MIN, MINS, and MINP), and those with $\Delta t = \delta$ (for EXACT, EXACTS, and EXACTP). We also note that the queries can be combined using logical operators (i.e., \emph{AND}, \emph{OR}, \emph{NOT}). Given that the queries return sets, this can be achieved by leveraging set operations such as intersection ($\cap$), union ($\cup$), and difference ($\setminus$).

\subsection{Time and Space Complexity Analysis}\label{sec:companalysis}

\begin{wraptable}[12]{R}{0.70\textwidth}
\centering
{\scriptsize{
\def\arraystretch{1.2} 
\setlength{\tabcolsep}{3pt} 
\begin{tabular}{c|c||c|c||c|c}
	
    \hline
    \multicolumn{2}{c||}{\textbf{Algorithm~1 (time)}}
	& \multicolumn{2}{c||}{\textbf{Algorithm~2 (time)}}
	& \multicolumn{2}{c}{\textbf{Data Structure (space)}}
    \\\hline
    
    \textbf{worst} & \textbf{average} & \textbf{worst} & \textbf{average} & \textbf{worst} & \textbf{average}\\
	$O({e^2} \cdot \log {e})$ & $O(e)$ & $O(e^2)$ & $O(\log {l})$ or $O(l)$ & $O(e^2)$ & $O(e)$\\\hline\hline
	
	\multicolumn{2}{c||}{\textbf{MR}}
	& \multicolumn{2}{c||}{\textbf{MR (time)}}
	& \multicolumn{2}{c}{\textbf{MR (space)}}
    \\\hline
    
    \textbf{worst} & \textbf{average} & \textbf{worst} & \textbf{average} & \textbf{worst} & \textbf{average}\\
	$N/A$ & $N/A$ & $O(e)$ & $O(e)$ & $O(e)$ & $O(e)$\\\hline\hline
	
	\multicolumn{2}{c||}{\textbf{MR with B+-Tree}}
	& \multicolumn{2}{c||}{\textbf{MR with B+-Tree(time)}}
	& \multicolumn{2}{c}{\textbf{MR with B+-Tree(space)}}
    \\\hline
    
    \textbf{worst} & \textbf{average} & \textbf{worst} & \textbf{average} & \textbf{worst} & \textbf{average}\\
	$O(e)$ & $O(e)$ & $O(e)$ &  $O(log_b~e  + (n\cdot~k))$ & $O(e)$ & $O(e)$\\\hline

	\end{tabular}
}}
  	\caption{Time and space complexity ($l$ -- log traces; $e$ -- log events; $n$ -- number of blocks; $k$ -- blocking factor).}\label{tab:complexity}
\end{wraptable}
The overarching goal of this study is the design of a data structure that can be used to check temporal compliance rules efficiently. Here, we provide the time complexity analysis for generating and accessing our proposed data structure -- referring to the operations described in Alg.~1 and Alg.~2 and the space complexity analysis for storing our proposed data structure. We compare, where applicable, our proposed data structure to using pattern matching on tuples in a relational database using the \emph{Match\_Recognize} SQL operator. The results are summarised in Table~\ref{tab:complexity}. 

\noindent\textbf{Data Structure Generation (Alg.~1).} To populate our data structure, we iterate one single time on the events recorded in the log; this operation is $O(e)$ -- where $e$ is the number of events in the log. For each event, we perform several operations. The first 14 operations are $O(1)$ (lines 15-28).~\footnote{We recall that adding/retrieving elements to/from maps with integer keys is $O(1)$.}
The nested iteration (lines 29-32) performs three operations as many times as the length of the processed partial trace, which grows by progressing the reading of the event log. Hence, the length of this nested iteration is $1, 2, 3, \dots, n$ -- where $n$ is the trace length.
The three operations have complexity of $O(1)$ (\emph{if} check); $O(1)$ (\emph{add} operation on map $\deltas$); and $O(\log {y})$ (\emph{add} operation on the red-black tree, where $y$ is the number of nodes in the tree).

Considering the worst scenario, we can assume that the nested iteration is $O(\frac{m+1}{2})$,~\footnote{Here, we are taking into account that the iterations increment linearly from $1$ to $m$.} where $m$ is the maximum trace length in the given log, and the nodes in the tree are $e^2$. Then, we obtain $O(e\cdot\frac{m+1}{2}\cdot\log {e^2}) = O({e} \cdot {m} \cdot \log {e})$. Although $m$ is not a constant, in the worst case its value is $e$ (i.e., the log is a single trace). We conclude that in the worst scenario, the time complexity for generating our data structure is $O({e^2} \cdot \log {e})$. However, on an average scenario (i.e., $m$ being the average trace length), we can reasonably replace $m$ with a large constant (e.g., $1000$). Given that the add operation on the red-black tree on an average scenario is $O(1)$, the resulting time complexity is $O(e)$. 

Coming to the MR solution, when not considering auxiliary indexes, no extra data structures are required. In the case of using indexing~\cite{indexAcceleratedPatternMatching21}, the index construction requires a linear scan on the input table (i.e., the log), which is always $O(e)$. 

\noindent\textbf{Data Structure Querying (Alg.~2).} For each query we want to resolve, we access one red-black tree at a specific node and traverse the tree towards the root, the leaves, or both (depending on the query) to collect all the case integer IDs that breach the given compliance rule. Accessing the tree at a given node has a complexity of $O(\log {y})$ -- where $y$ is the number of nodes in the tree. Traversing the tree by reading all the nodes (to collect the case integer IDs) requires $y$ iterations in the worst scenario. 
In addition, when resolving queries of type MAXP, MINP, and EXACTP, we need to iterate over two additional trees (lines 25 and 30). 
Considering this, in the worst scenario, the time complexity to answer one query is $O( 3y + \log {y})$, maximizing $y$ as $e^2$, we obtain $O( 3e^2 + \log {e^2}) = O(e^2)$. However, this is a truly pessimistic scenario, far from being realistic, as also our extensive evaluation shows. It is unlikely -- if possible at all -- that one red-black tree would have $e^2$ nodes, given that one red-black tree captures the $\Delta t$ of a pair of activities. A realistic (and average) assumption is that the number of nodes of one red-black tree is equal to the traces recorded in the event log multiplied by a factor ($K$) that would depend on the maximum number of observation of each activity within a trace. Under this assumption, maximizing $K$ with a large number (e.g., $100$ or $1000$), we would obtain $O(l + \log {l}) = O(l)$, where $l$ is the number of traces in the log. Furthermore, we note that the iterations over the subtree nodes (lines 10, 16, 21) are merely required to provide the output to the user, rather than to answer the queries. The detection of the subtrees identifies all the breaching cases. From this perspective, the query resolution can be considered $O(\log {l})$ for the queries MAX, MIN, EXACT, MAXS, MINS, and EXACTS. Our empirical evaluation supports this result.

Finding patterns using MR requires a linear scan on the input table when no auxiliary indexes are available, as we have to check each record against the pattern. When using indexes, i.e., a B+-Tree, we require index access for each symbol in the pattern. In our case, we have two symbols $A$ and $B$; hence, each access is $O(log_b~e)$ to the index, where $b$ is the number of keys per B+-tree block. Each access results in $n$ disk blocks, where each block holds $k$ tuples.~\footnote{$k$ is the blocking factor. That is how many events are stored per block.} Then, we have to scan $n\cdot~k$ tuples for matching the pattern. In total, we have $O(2(log_b~e  + n\cdot~k))$ operations to perform. In the worst scenario, when the selectivity of $A$, $\mu(A)$, is very low, i.e., events of $A$ occur frequently, we tend to visit the same number of blocks as there are in the table. That is $\lim_{\mu \to 1}~n\cdot~k = e$.\footnote{If the selectivity of $A$ is high enough, the query optimizer may bypass index access and resort to a table scan.} In total, we have $O(2(log_b~e  + e))$ which is $O(e)$. In the average case, when the selectivity of the symbols is low, we tend to retrieve references to a small number of blocks $n$. On average the complexity is  $O(log_b~e  + (n\cdot~k))$

\noindent\textbf{Data Structure Space Complexity.} We calculate our space complexity in terms of memory elements, considering the worst scenario. Our data structure includes 2 integer counters and 6 maps (see Table~\ref{tab:datastruct}). $C_c$ and $C_a$ are constant. $\cIDs$ has size always equal to the number of traces in the event log ($l$). $\aIDs$ has size always equal to the number of distinct activities ($\sigma$). $\obs$ has size equal to $\sigma \cdot l$. $\activities$ and $\timestamps$ have equal size, which is always the total number of events ($e$) recorded in the log. $\deltas$ has the most variable size, in the worst scenario (as discussed above), we have $\sigma^2$ elements in $\deltas$, each element is a red-black tree with (at most) $e^2$ elements. Hence, in the worst scenario, the space complexity of our data structure is
$O\left(l + \sigma + \sigma \cdot l + 2e + \sigma^2 \cdot e^2 \right)$. 
However, we can reasonably replace $\sigma$ with a large constant (e.g., $100$ or $1000$), obtaining $O\left(l + e + e^2 \right) = O\left(e^2 \right)$. However, if we consider an average scenario, as we mentioned above, $e^2$ is likely to be approximately $l$,
obtaining $O\left(l + e + l \right) = O(e)$. For the case of MR with indexing, we need space to store the each event of the log, which is $O(e)$, and an auxiliary index (for the B+-Tree), that is also $O(e)$.

\subsection{Online Iterative Update and Optimizations}\label{sec:itupdate}

When our proposed data structure (as described in Table~\ref{tab:datastruct}) is kept in memory or disk, it is possible to update it by re-executing Alg.~1 -- clearly, skipping lines 1-8, which are initialization operations. This follows by design. Because Alg.~1 receives as input a sequence of events, either as a log or as a stream (i.e., one single event at a time), and it keeps track of the past events via the maps $\activities$ and $\timestamps$. In fact, the only purpose of these two maps is to allow for a continuous update of the data structure. We note that the maps $\activities$ and $\timestamps$ can be expensive (memory-wise) to maintain; if a continuous update is not required, they can be discarded after executing Alg.~1.

To simplify the description of our data structure, we did not discuss the scenario where the events in the input log capture activity lifecycle stages, e.g., the start and the completion of an activity. To process these events, we have one of the following two options. Option-1; we retain all the activity lifecycle stages as different activities, e.g., we would have activity $Y_s$ (start of $Y$) and activity $Y_c$ (completion of $Y$). Option-2; we filter the activity pairs that we store in the map $\deltas$.
Option-1 provides maximum flexibility at the cost of additional memory space, and it does not alter Alg.~1 and Alg.~2. It only requires that queries' source and target activities include their lifecycle stages. Option-2 requires a minor alteration to Alg.~1. Precisely, within the iteration at line 29, an overarching check on the pair $(A_k, a)$ is to establish whether that type of pair (depending on the lifecycle stage that $A_k$ and $a$ refer to) should be stored in the map $\deltas$ or not. If not, the iteration would proceed to the next element; if yes, the iteration would unfold as described in lines 30-32.

An important optimization is to alter the red-black tree nodes from a pair ($\Delta t$, $c$) to a pair ($\Delta t$, $\hat{C}$), where $\hat{C}$ is a set of case integer IDs as opposed to a single integer ID.  The order of the red-black tree nodes would be determined only by $\Delta t$. Consequently, all the pairs/nodes ($\Delta t$, $c$) having the same $\Delta t$ would be grouped into a single pair/node as ($\Delta t$, $\hat{C}$). Such variation reduces the number of nodes in each tree, leading to better time complexity.


\section{Evaluation}\label{sec:eval}
\vspace{-3mm}

We have implemented our data structure and its query engine as a Java command-line tool.\footnote{Available at: \scriptsize{\url{https://doi.org/10.6084/m9.figshare.17090006}}}
We empirically verified the efficiency of our approach by running three experiments. Here, we describe the dataset, the setup, and the results of our evaluation.

\noindent\textbf{Datasets and Setup.} For our experiments, we selected a collection of 12 public event logs from a benchmark of automated process discovery algorithms~\cite{augusto2018automated}
\footnote{\scriptsize{\url{https://doi.org/10.4121/uuid:adc42403-9a38-48dc-9f0a-a0a49bfb6371}}}, and then we extended this collection with eight additional event logs, sourced from the Business Process Intelligence Challenges of 2019~\footnote{\scriptsize{\url{https://data.4tu.nl/articles/dataset/BPI_Challenge_2019/12715853/1}}} (three out of eight) and 2020~\footnote{\scriptsize{\url{https://data.4tu.nl/collections/BPI_Challenge_2020/5065541/1}}}(five out of eight). Table~\ref{tab:logs} provides an overview of the logs features, highlighting the heterogeneous nature of the dataset. We used these logs to run the following three experiments. 

\begin{wraptable}[17]{R}{0.61\textwidth}
 \vspace{-2mm}
\centering
{\scriptsize{
\begin{tabular}{cl|cc|cc|ccc}

    \hline
    \multicolumn{2}{c|}{\textbf{Logs}} 
    & \multicolumn{2}{c|}{\textbf{Traces}} 
    & \multicolumn{2}{c|}{\textbf{Events}} 
    & \multicolumn{3}{c}{\textbf{Trace Length}} \\\hline
    
    \textbf{ID}
    & \textbf{Label}
    & \textbf{total}
    & \textbf{unique (\%)} 
    & \textbf{total}
    & \textbf{unique (\#)} 
    & \textbf{min} 
    & \textbf{avg} 
    & \textbf{max} \\\hline
    
    1 & \textbf{BPIC12} & 13087 & 33.4  & 262200 & 24    & 3     & 20    & 175 \\
    2 & \textbf{BPIC13cp} & 1487  & 12.3  & 6660  & 4     & 1     & 4     & 35 \\
    3 & \textbf{BPIC13inc} & 7554  & 20.0  & 65533 & 4     & 1     & 9     & 123 \\
    4 & \textbf{BPIC14} & 41353 & 36.1  & 369485 & 9     & 3     & 9     & 167 \\
    5 & \textbf{BPIC15f1} & 902   & 32.7  & 21656 & 70    & 5     & 24    & 50 \\
    6 & \textbf{BPIC15f2} & 681   & 61.7  & 24678 & 82    & 4     & 36    & 63 \\
    7 & \textbf{BPIC15f3} & 1369  & 60.3  & 43786 & 62    & 4     & 32    & 54 \\
    8 & \textbf{BPIC15f4} & 860   & 52.4  & 29403 & 65    & 5     & 34    & 54 \\
    9 & \textbf{BPIC15f5} & 975   & 45.7  & 30030 & 74    & 4     & 31    & 61 \\
    10 & \textbf{BPIC17} & 21861 & 40.1  & 714198 & 18    & 11    & 33    & 113 \\
    11 & \textbf{RTFMP} & 150370 & 0.2   & 561470 & 11    & 2     & 4     & 20 \\
    12 & \textbf{SEPSIS} & 1050  & 80.6  & 15214 & 16    & 3     & 14    & 185 \\
    13 & \textbf{BPIC19a} & 15129 & 20.9  & 283407 & 5     & 1     & 19    & 794 \\
    14 & \textbf{BPIC19b} & 220810 & 1.2   & 979942 & 8     & 1     & 4     & 179 \\
    15 & \textbf{BPIC19c} & 1027  & 10.1  & 5038  & 4     & 2     & 5     & 19 \\
    16 & \textbf{BPIC20a} & 10500 & 0.9   & 56437 & 17    & 1     & 5     & 24 \\
    17 & \textbf{BPIC20b} & 6449  & 11.7  & 72151 & 34    & 3     & 11    & 7 \\
    18 & \textbf{BPIC20c} & 7065  & 20.9  & 86581 & 51    & 3     & 12    & 90 \\
    19 & \textbf{BPIC20d} & 2099  & 9.6   & 18246 & 29    & 1     & 9     & 21 \\
    20 & \textbf{BPIC20e} & 6886  & 1.3   & 36796 & 19    & 1     & 5     & 20 \\\hline
    
	\end{tabular}
}}
  	\caption{Logs features.}\label{tab:logs}
\end{wraptable}%

\noindent\textbf{Exp-1.} Firstly, we performed a monolithic reading of each of the logs to populate our data structure, with subsequent querying -- using 5, 10, 50, and 100 mixed queries. In this experiment, we measured the execution times to process each log and to resolve the queries, comparing their execution times and the number of average hits.

\noindent\textbf{Exp-2.} Secondly, we divided each log into five partitions. Each partition had an equal number of traces. We processed one partition at a time (i.e., recurrently running Algorithm~1) while saving/loading our data structure to/from disk in the form of a CSV file. We compared the execution times of this experiment with those of Exp-1.

\noindent\textbf{Exp-3.} Thirdly, we selected the six largest logs in the dataset (by the total number of events), i.e., BPIC12, BPIC14, BPIC17, RTFMP, BPIC19a, BPIC19b, and we divided them into five partitions. This time, each partition captured a timeframe equal to approximately $1/5$ of the total. For instance, the RTFMP log records events from $01-01-2000$ to $18-06-2013$. So, each of its partitions was approximately $32$ months and a half. Each log partition had several fragmented traces, i.e., traces that began/ended in a previous/successive partition. This experiment allowed us to assess how efficiently our data structure can be updated when traces are fragmented across logs. We compared the results of this experiment with those of Exp-1 and Exp-2.

All the experiments were run on an Intel Core i7-8565U@1.80GHz with 32GB RAM running Windows 10 Pro (64-bit) and Java 11, with 14GB RAM of maximum memory allocation for the JVM and 10GB maximum stack size.

\begin{figure}
\vspace{-5mm}
	\centering
	\subfloat[Time complexity vs Execution time (ns)]{
		\includegraphics[width=0.47\textwidth]{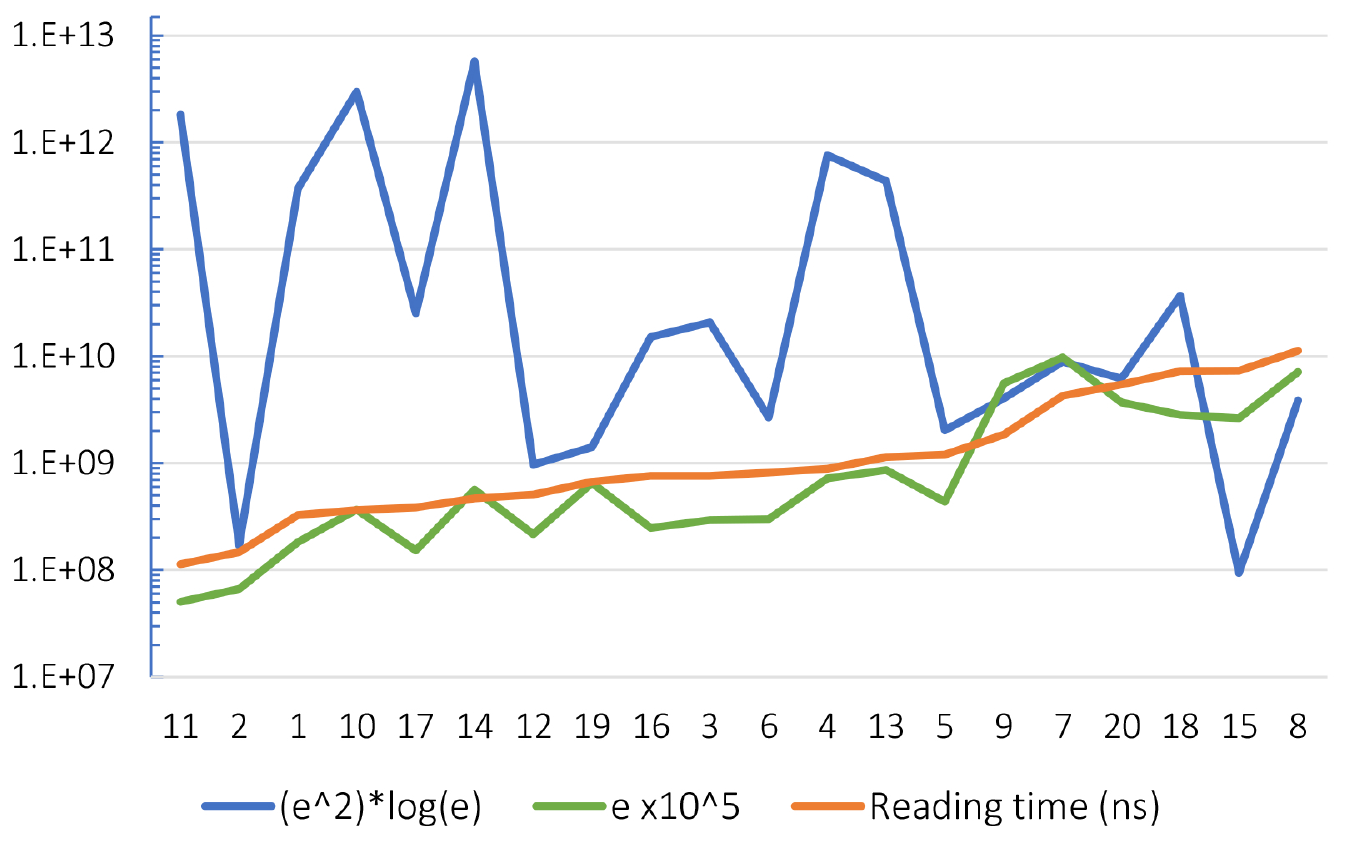}\label{fig:timecomp}
	}
	\hfill
	\subfloat[Execution times (s)]{
		\includegraphics[width=0.47\textwidth]{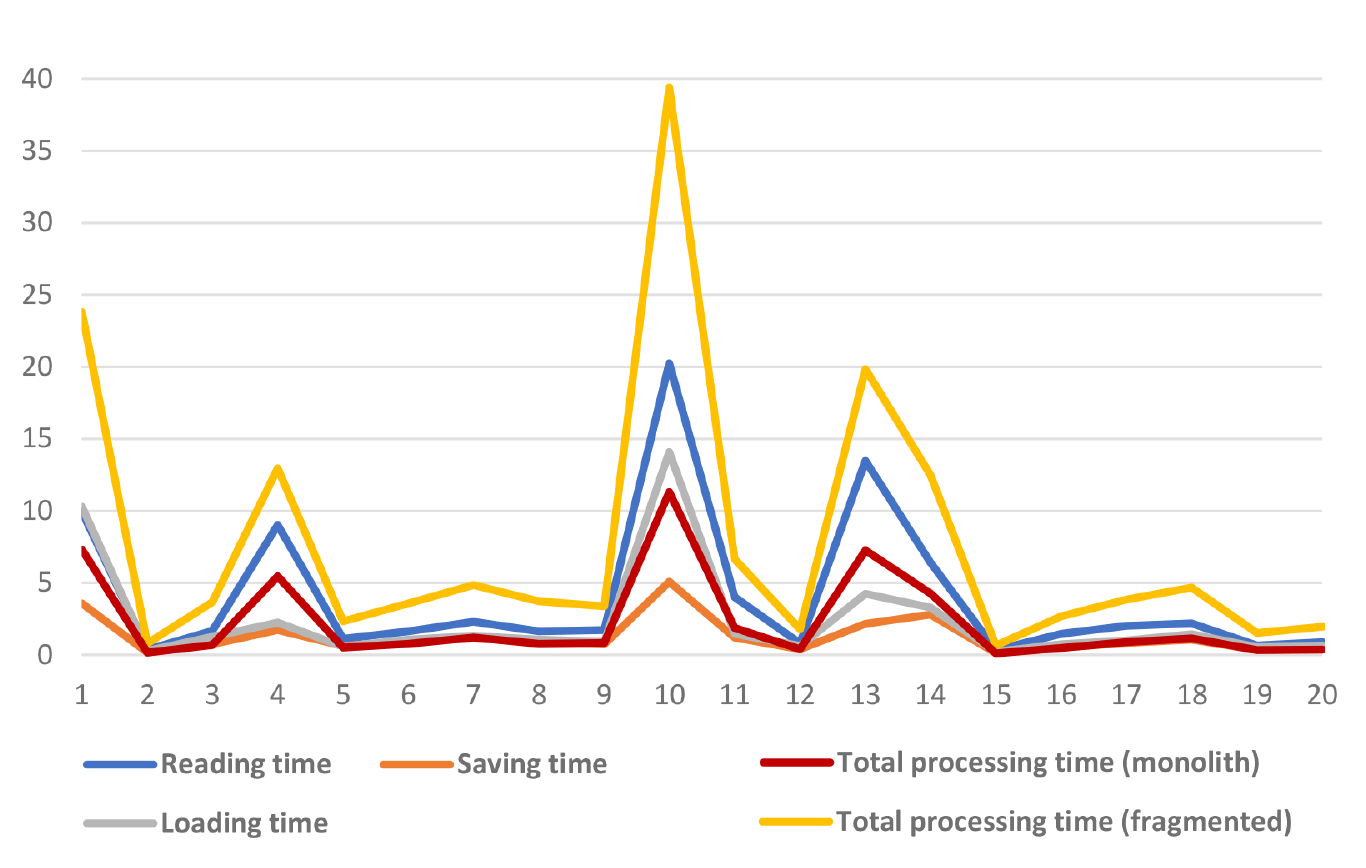}\label{fig:exp12-times}
	}
\caption{Exp-1 and Exp-2 results.}
\label{fig:exp1}
\end{figure}

\noindent\textbf{Results and Discussion.} Table~\ref{tab:queries-time} and Fig.~\ref{fig:exp1} report the results of Exp-1. The graph in Fig.~\ref{fig:timecomp} (x-axis = log IDs, see Table~\ref{tab:logs}) shows that the time complexity analysis we reported in Section~\ref{sec:companalysis} for Algorithm~1 is reflected in our empirical evaluation, with the real execution time (in nanoseconds) having a trend extremely close to $O(e)$ (average scenario time complexity), and in rare occasions to $O(e^2 \cdot \log e)$ (worst scenario time complexity). Considering Fig.~\ref{fig:exp12-times}, we note that the time required to generate our data structure is close to the second for the vast majority of the logs, except for those logs recording a large number of events ($>200k$) -- where we note clear peaks in execution times. Nevertheless, even in the worst case (BPIC17, 700k+ events), the execution time is $11.3$s, which is reasonable for an algorithm that is not supposed to run in real-time, as opposed to the querying algorithm (Algorithm~2).
Table~\ref{tab:queries-time} reports the execution times for querying our data structure and the average hits. These results are challenging to reconcile with our theoretical complexity analysis of the querying algorithm. The reason is the difficulty of estimating the number of unique trees and their nodes in our data structure and generalising this for a range of queries that may largely differ in terms of hits. Nonetheless, the execution times are positively remarkable, with querying times consistently below the second -- even when detecting millions of violating cases.

\begin{table}[h!]
{\scriptsize{
  \centering
\def\arraystretch{1.2} 
\setlength{\tabcolsep}{3pt} 
    \begin{tabular}{l|r|rr|rr|rrrr|rrrr}
    
    \hline
    \textbf{Log} 
    &       
    & \multicolumn{2}{c|}{\textbf{Events}} 
    & \multicolumn{2}{c|}{\textbf{Trees}}   
    & \multicolumn{4}{c|}{\textbf{Cumulative time (ms)}} 
    & \multicolumn{4}{c}{\textbf{Cumulative hits}} \\\cline{3-14}
    
    \textbf{ID}
    & \multicolumn{1}{c|}{\textbf{Traces}} 
    & \multicolumn{1}{c}{\textbf{Total}} 
    & \multicolumn{1}{c|}{\textbf{Unique}} 
    & \multicolumn{1}{c}{\textbf{Unique}} 
    & \multicolumn{1}{c|}{\textbf{Nodes}} 
    & \multicolumn{1}{c}{\textbf{5}} 
    & \multicolumn{1}{c}{\textbf{10}} 
    & \multicolumn{1}{c}{\textbf{50}} 
    & \multicolumn{1}{c|}{\textbf{100}} 
    & \multicolumn{1}{c}{\textbf{5}} 
    & \multicolumn{1}{c}{\textbf{10}} 
    & \multicolumn{1}{c}{\textbf{50}} 
    & \multicolumn{1}{c}{\textbf{100}} \\\hline

    1     & 13087 & 262200 & 24    & 388   & 5360356 & 15    & 47    & 137   & 163   & 21154 & 45073 & 302695 & 617769 \\
    2     & 1487  & 6660  & 4     & 18    & 25720 & $< 1$     & 15    & 24    & 47    & 1614  & 7253  & 27926 & 67161 \\
    3     & 7554  & 65533 & 4     & 18    & 536626 & 16    & 50    & 114   & 112   & 16125 & 23372 & 124693 & 266161 \\
    4     & 41353 & 369485 & 9     & 63    & 3238883 & 16    & 80    & 237   & 312   & 79260 & 145958 & 782717 & 1646844 \\
    5     & 902   & 21656 & 70    & 2154  & 363491 & $< 1$     & 4     & 12    & $< 1$     & 1831  & 2108  & 17273 & 36087 \\
    6     & 681   & 24678 & 82    & 3271  & 572478 & $< 1$     & $< 1$     & 11    & 16    & 2019  & 2943  & 18019 & 37948 \\
    7     & 1369  & 43786 & 62    & 1991  & 823628 & 17    & 5     & 16    & 7     & 3312  & 4592  & 31896 & 67093 \\
    8     & 860   & 29403 & 65    & 2013  & 596658 & $< 1$     & $< 1$     & 12    & 7     & 1252  & 2020  & 16564 & 35574 \\
    9     & 975   & 30030 & 74    & 2629  & 641663 & $< 1$     & $< 1$     & 15    & 16    & 2865  & 3782  & 25264 & 52957 \\
    10    & 21861 & 714198 & 18    & 187   & 12770140 & 15    & 52    & 115   & 146   & 41807 & 72492 & 435858 & 975939 \\
    11    & 150370 & 561470 & 11    & 95    & 1531445 & $< 1$     & 79    & 244   & 98    & 8767  & 25266 & 105002 & 217137 \\
    12    & 1050  & 15214 & 16    & 179   & 186904 & $< 1$     & 21    & 38    & 48    & 3217  & 6108  & 33695 & 68326 \\
    13    & 15129 & 283407 & 5     & 25    & 19761885 & 21    & 62    & 184   & 198   & 18810 & 50280 & 229339 & 495281 \\
    14    & 220810 & 979942 & 8     & 56    & 2994505 & 31    & 133   & 632   & 781   & 143805 & 311215 & 1580527 & 2916536 \\
    15    & 1027  & 5038  & 4     & 17    & 21876 & 4     & 9     & 24    & 32    & 1776  & 3134  & 16025 & 35393 \\
    16    & 10500 & 56437 & 17    & 160   & 191490 & 11    & 38    & 96    & 147   & 23265 & 38433 & 180412 & 424505 \\
    17    & 6449  & 72151 & 34    & 693   & 463906 & 6     & 20    & 46    & 64    & 11048 & 31515 & 108079 & 266942 \\
    18    & 7065  & 86581 & 51    & 1465  & 683992 & $< 1$     & 21    & 55    & 84    & 12900 & 37567 & 129576 & 317701 \\
    19    & 2099  & 18246 & 29    & 477   & 93758 & $< 1$     & 10    & 29    & 36    & 4658  & 9266  & 43485 & 101924 \\
    20    & 6886  & 36796 & 19    & 167   & 124123 & $< 1$     & 14    & 48    & 62    & 546   & 4473  & 46188 & 109545 \\
\hline
    \end{tabular}%
    }}
  \caption{Exp-1 queries results (times in milliseconds)}
  \label{tab:queries-time}%
\end{table}%

Considering the results of Exp-2, in Fig.~\ref{fig:exp12-times}, we note that the execution time overhead of \emph{reading} (i.e., running Algorithm~1) fragmented logs is little or negligible, except for three logs (BPIC17, BPIC19a, and BPIC19b). However, as expected, \emph{saving} our data structure on the disk (after processing a log fragment) and \emph{loading} it again (before processing a new log fragment) have an impact on the total execution time. This ultimately leads to a clear overhead, as shown by the yellow and red trends in Fig.~\ref{fig:exp12-times}.
In practice, it would be better to process a log in one-pass rather than sequentially process its fragments. Though, we remark that alternative storing and access methods could be devised to optimize such procedures and that processing in parallel log fragments could also dramatically improve the results -- we leave this to future research endeavours.

Lastly, Table~\ref{tab:exp3} shows the results of Exp-3. When comparing the \emph{reading}, \emph{saving}, and \emph{loading} times of Exp-3 against those of Exp-2, the former are most of the time better than or in-line with the latter (15 times out of 24). This highlights that processing log fragments with incomplete traces do not substantially affect the efficiency of updating our data structure recurrently. Most of the execution time overhead, when compared to the processing of full logs (i.e., Exp-1), is due to the \emph{saving} and \emph{loading} operations.
\begin{table}[htbp]
{\scriptsize{
  \centering
\def\arraystretch{1.2} 
\setlength{\tabcolsep}{3pt} 
    \begin{tabular}{l|r|rrrr|rrrr|r}
    
           \hline
          & \multicolumn{5}{c|}{\textbf{Fragmented by Timeframe (Exp-3)}} 
          & \multicolumn{4}{c|}{\textbf{Fragmented by Traces (Exp-2)}} 
          & \textbf{Monolith (Exp-1)} \\\cline{2-11}
          
            \textbf{Logs}
          & \textbf{Frag. traces} 
          & \textbf{R} 
          & \textbf{S} 
          & \textbf{L} 
          & \textbf{Tot.} 
          & \textbf{R} 
          & \textbf{S} 
          & \textbf{L} 
          & \textbf{Tot.} 
          & \textbf{Tot.}
          \\\hline
          
    \textbf{BPIC12} & 2858  & 6.9   & 2.8   & 5.8   & 15.5 & 10.0  & 3.6   & 10.3  & 23.8  & 7.3 \\
    \textbf{BPIC14} & 2371  & 7.5   & 1.8   & 2.2   & 11.5 & 9.0   & 1.7   & 2.2   & 13.0  & 5.5 \\
    \textbf{BPIC17} & 3925  & 22.1  & 5.1   & 14.1  & 41.3  & 20.3  & 5.1   & 14.1  & 39.4 & 11.3 \\
    \textbf{RTFMP} & 42604 & 3.9   & 2.4   & 2.7   & 9.0   & 4.0   & 1.2   & 1.5   & 6.7 & 1.9 \\
    \textbf{BPIC19a} & 11062 & 11.8  & 1.6   & 2.3   & 15.8 & 13.5  & 2.1   & 4.2   & 19.9  & 7.2 \\
    \textbf{BPIC19b} & 173395 & 8.4   & 3.4   & 6.9   & 18.7  & 6.4   & 2.8   & 3.3   & 12.5 & 4.3 \\\hline
    \end{tabular}%
    }}
  
  \caption{Exp-3 results. Reading (R), saving (S), and loading (L) execution times (s).}\label{tab:exp3}%
\end{table}%
\section{Conclusion}\label{sec:conclusion}

In this paper, we propose an approach to efficiently check the violation for a class of temporal compliance rules against event logs. Specifically, the approach is sublinear query time on the number of traces in the log (input) and linear on the number of violating traces (output). This efficiency is achieved by constructing an index-based data structure from the event log, which compactly captures the temporal relations between activity instances. 
The experimental evaluation is in line with the theoretical bounds on execution time, showing that our approach can be used to check a wide range of temporal compliance rules in sub-second time on event logs with up to a million events, and without having to fix the set of compliance rules in advance (i.e.\ interactive use).

There are several directions for future work. First, to cope with the growth of event logs, the index construction could be parallelized following a MapReduce style. Second, we plan to index more event attributes (e.g.\ resources) to cover more compliance patterns. Third, we plan to conduct further experiments on larger event logs and against baselines such as time-enhanced implementations of the Match\_Recognize operator.

\medskip\noindent\textbf{Acknowledgments.} Work funded by European Regional Development Funds (Mobilitas Plus programme grant MOBTT75) and Estonian Research Council (PRG1226).

\bibliographystyle{plain}
\bibliography{lit}

\end{document}